\documentclass[10pt,pre,superscriptaddress,twocolumn,showpacs,floatfix]{revtex4-1}
\usepackage[utf8x]{inputenc} 
\usepackage{mathtools}
\usepackage{amsfonts}
\usepackage{amsmath}
\usepackage{amsthm}
\usepackage{amssymb}
\usepackage{graphicx}
\usepackage{rotating}
\newcommand{\IFUFG}{Instituto de F{\'{\i}}sica, Universidade Federal de 
Goi{\'a}s, Av. Esperan\c{c}a s/n, 74.690-900, Goi{\^a}nia, GO, Brazil}
\newcommand{\IFMT}{Instituto Federal do Mato Grosso - Campus C{\'a}ceres, 
Av. dos Ramires s/n, 78200-000, C{\'a}ceres, MT, Brazil}
\begin{document}

\title{Critical behavior of the spin-$1/2$ Baxter-Wu model: Entropic 
sampling  simulations}

\author{L. N. Jorge}
\affiliation{\IFUFG}
\affiliation{\IFMT}
\author{L. S. Ferreira}
\affiliation{\IFUFG}
\author{S. A. Le\~ao}
\affiliation{\IFUFG}
\author{A. A. Caparica}
\affiliation{\IFUFG}
\email{caparica@ufg.br}

\begin{abstract}
In this work we use a refined entropic sampling technique based on the 
Wang-Landau method to study the spin-$1/2$ Baxter-Wu model.  The static 
critical exponents were determined as $\alpha=0.6545(68)$, $\beta=0.0818(30)$, 
$\gamma=1.18193(77)$, and $\nu=0.66341(47)$. The estimate for the critical 
temperature was $T_c=2.269194(45)$. We compare the present results with 
those obtained from other well established approaches and we find a very good
closeness with the exact values, besides the high precision reached for the 
critical temperature. We also calculate the coefficients $a$ and $b$ for the 
divergence of the microcanonical inverse temperature at the ground state 
achieving an excellent agreement in comparison with the simulation estimates. 
\end{abstract}

\maketitle

\section{Introduction}

In the last decade of the $20^{th}$ century, many researchers contributed to 
the construction of a new way of performing Monte Carlo simulations without 
fixing the temperature as it is done in the Metropolis method. Good reasons for 
this trend were to avoid the critical slowing down in second order phase 
transitions and to overcome the tunneling barrier between coexisting phases at 
the transition temperature in first order phase transitions. To mention a few 
efforts in this direction, there is the multicanonical 
method\cite{berg1,berg2,berg3,janke}, the entropic sampling method\cite{lee} 
(today it is usual to refer to all the methods that estimate
$\ln g(E)$, where $g(E)$ is the density of states, as entropic sampling 
methods), and the broad histogram algorithm\cite{broad1,broad2,broad3} among 
others. The culmination of this sequence of works was the Wang-Landau 
algorithm\cite{WangLandau2001} which yielded very impressive results in 
estimating the density of states. Zhou and Bhatt\cite{ZhouBhatt2005} 
demonstrated the convergence of the method, on the other hand Belardinelli 
and Pereyra\cite{belardinelli} have shown that the error saturates much earlier 
than expected in the original paper. Caparica and Cunha-Netto\cite{caparica} 
proposed some easily implementable changes to the method in order to improve 
accuracy, such as adopting the Monte Carlo step for updating the density of 
states and excluding the initial WL simulation levels from the microcanonical 
averages.  In a later paper Caparica\cite{caparica3} proposed a criterion for 
halting the simulations and demonstrated that the convergence of Ref. 
\cite{ZhouBhatt2005} is not towards the correct value, but tends to a value
within a Gaussian distribution. Furthermore it was demonstrated in that 
work that even taking into account all the improvements proposed in
Ref.\cite{caparica} the final results still fall into a Gaussian distribution 
and to overcome these difficulties one should perform at least ten independent 
sets of finite-size scaling simulations.

In the present work we carry out a study of the Baxter-Wu model applying the 
improved entropic method following Refs. 
\cite{WangLandau2001,caparica,caparica3} combining finite-size scaling and 
cumulant methods.

The Baxter-Wu (BW) model although less widespread than the Ising model 
can also be considered a benchmark in statistical physics. It exhibits a 
second order phase transition and there are exact solutions
for the critical temperature and critical exponents. The model was 
solved by R.J. Baxter and F.Y. Wu\cite{bw1,bw2,bw3}. It consists of a
two-dimensional triangular lattice with $\sigma_i =\pm1$ spins located in
the vertices of the triangles and interacting via a three spin interaction
and having the same critical temperature of the Ising model and 
critical exponents $\alpha=\nu=2/3$, $\beta=1/12$ and $\gamma=7/6$, belonging
therefore to the same universality class of the $q=4$ Potts model. This
should be intuitively expected, since the BW model has the same symmetry and 
the same degree of degeneracy in the ground state of the $q=4$ Potts model with
four different configurations. Therefore, the BW model is a rich ground for 
performing and testing simulational methods.

The outline of this paper is as follows: In section II we define the model. In 
section III we describe the simulation procedure. In section IV we present the 
finite-size scaling analysis. The results are discussed in section V and we 
devote section VI to the summary and concluding remarks. 

\section{Baxter-Wu Model}

The BW model was initially proposed by D. W. Wood and H. P. 
Griffiths\cite{wood} in 1972. This model is defined in a triangular 
two-dimensional lattice, such that a three-spin interaction is given by the 
Hamiltonian,
\begin{equation}
  {\cal H} = -J \sum_{<i,j,k>} \sigma_i  \sigma_j \sigma_k,
\label{hamiltonian}  
\end{equation}
where the spin variables are located at the vertices of the lattice and 
take the values $\sigma_i=\pm1$, $J$ is the coupling constant that defines the 
energy scale and the sum extends over all triangular faces of the lattice.
One of the reasons that motivated the construction of this model was the 
elaboration of a magnetic model in which there is no symmetry by inversion of 
spins and that exhibits an order-disorder transition. 

\begin{figure}[!h]
\begin{center}
\includegraphics[width=.95\linewidth]{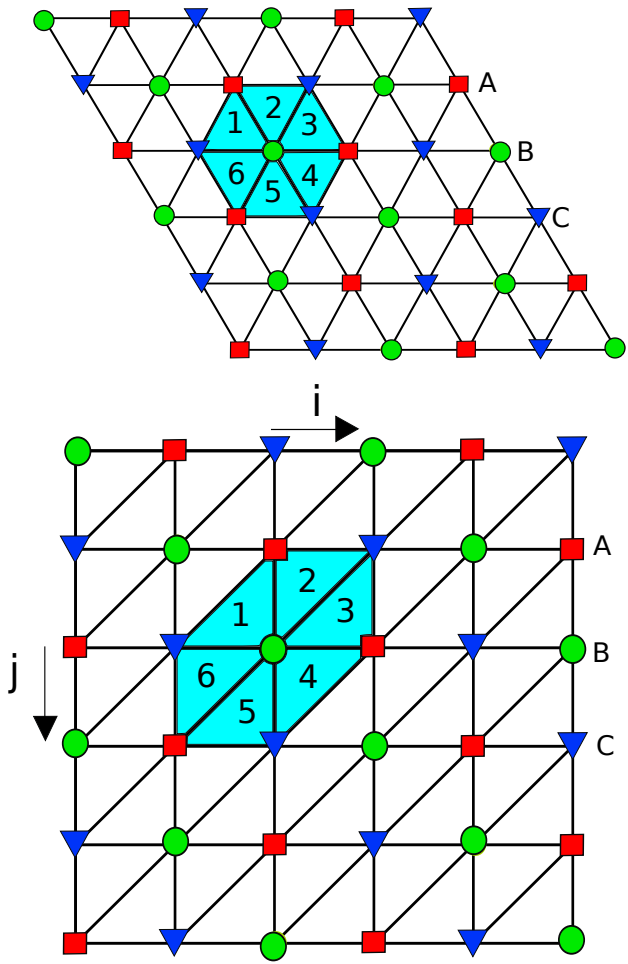}
\end{center}
\vspace{-.5cm}
\caption{(color online). Top: Representation of the Baxter-Wu model as 
a superposition of three sub-lattices on a triangular lattice.  Bottom:
Transposition of the triangular lattice to a square lattice.} \label{baxterwu}
\end{figure}

The triangular lattice can be divided into three sub-lattices \textit{A, B, C}, 
as represented in top of Fig.\ref{baxterwu}, so that any triangular face 
(\textit{i, j, k}) contains one site of type \textit{A}, one of type 
\textit{B}, and one of type \textit{C}.

The model exhibits four distinct ground state configurations, namely one with
all spins positive and still other three where 
the spins of two sub-lattices are negated successively.

The configurations on the triangular lattice may be transposed to isomorphic 
configurations on a square lattice as shown in bottom of Fig.\ref{baxterwu}.
One can therefore deal with the model considering spins $\sigma_i\equiv s_{ij}$
on a square lattice, using periodic boundary conditions. Each spin is 
surrounded by six triangular faces, as shown in Fig.\ref{baxterwu}. If one 
runs all the spins of the configuration counting six faces for each spin, each 
triangular face will be considered three times. Therefore, the energy of a 
given configuration of lattice size $L$ defined by the Hamiltonian 
\eqref{hamiltonian} can be calculated in the square lattice as

\begin{align}
  E =&-\frac{J}{3} \sum_{i=1}^{L}\sum_{j=1}^{L}  
s_{i,j}(s_{i-1,j}s_{i,j-1}+s_{i,j-1}s_{i+1,j-1} \notag \\  
&+s_{i+1,j-1}s_{i+1,j}+s_{i+1,j}s_{i,j+1}
+s_{i,j+1}s_{i-1,j+1} \notag \\
& +s_{i-1,j+1}s_{i-1,j}) \label{energy}
\end{align}

Our simulations were performed using this square lattice scheme.

\section{Entropic Sampling Simulations}

The seminal idea of the Wang-Landau method \cite{WangLandau2001} is that a 
random walk in the energy space with a probability proportional to the inverse 
of the density of states generates a flat histogram for the energy distribution. 
In fact one estimates $S(E)\equiv\ln g(E)$, because estimating the density of 
states $g(E)$ would produce huge numbers. The simulation begins setting 
initially $S(E)=0$ for all energy levels. The random walk in the energy space 
runs through the energy levels with a probability $p(E\rightarrow 
E')=\min(\frac{g(E)}{g(E')},1)$, where $E$ and $E'$ are the energies of the 
current and the next possible configurations, respectively. If a configuration 
is accepted we update $H(E')\rightarrow H(E')+1$ and $S(E')\rightarrow 
S(E')+F_{i}$, where $F_{i}=\ln f_{i}$, $f_{0}\equiv e=2.71828...$ and 
$f_{i+1}=\sqrt{f_{i}}$, where $f_{i}$ is defined in the method as the 
modification factor. After a number of Monte Carlo steps (MCS) we check the 
histogram for flatness and usually we consider the histogram flat if 
$H(E)>0.8\langle H \rangle$, for all energies, where $\langle H \rangle$ is an 
average over the energies. If the flatness condition is satisfied, the 
modification factor is updated to a finer one and the histogram is reseted 
$H(E)=0$.

Recently some small changes in this procedure were proposed 
\cite{caparica,caparica3,caparica1,caparica2}. Namely (i) the density of states 
should be updated only after each Monte Carlo sweep\cite{mcs}, since it 
avoids taking into account highly correlated configurations when 
estimating the density of states; (ii) the simulations should be carried out 
only up to $\ln f=\ln f_{final}$ defined by the canonical averages during the 
simulations. In this case we save CPU time, discarding superfluous long 
simulations; and (iii) the initial WL levels should be neglected before 
accumulating the microcanonical averages, because the configurations in 
the beginning of the simulations do not match those of maximum entropy. The
adoption of these easily implementable modifications leads to more accurate 
results and saves computational time. They investigated the behavior of the 
maxima of the specific heat
\begin{equation}\label{cv}
 C(T)=\langle(E-\langle E\rangle)^2\rangle/T^2
\end{equation}
and the magnetic susceptibility
\begin{equation}\label{ki}
 \chi(T)=L^2\langle(m-\langle m\rangle)^2\rangle/T,
\end{equation}
where $E$ is the energy of a given configuration and $m$ is the corresponding 
magnetization per spin, during the WL sampling for the Ising model on a square 
lattice. They demonstrated that a substantial part of the conventional 
Wang-Landau simulation is not necessary because the error saturates. They 
observed that in general no single simulation run converges to the real value 
of any physical quantity, but to a particular value of a Gaussian distribution 
of results around the true value. The saturation of the error matches with 
the convergence to this value. Simulations beyond this limit leads to negligible
variations in the canonical averages of all thermodynamic variables.

Another innovation proposed in Ref.\cite{caparica3} is a criterion for halting 
the simulations. During the WL sampling of a given model, beginning from 
$f_{5}$, one calculates the temperature of the peak of the specific heat 
defined in Eq. \eqref{cv} using the current $g(E)$ and from then on this 
mean value is updated whenever the histogram is checked for flatness. If the 
histogram is considered flat, we save the value of the temperature $T_c(0)$ of 
the peak of the specific heat. During the simulations with the new 
modification factor the temperature of the peak of the specific heat $T_c(t)$ 
continues being calculated when the histogram is checked for flatness along with
the checking parameter

\begin{equation}\label{eps}
 \varepsilon=|T_c(t)-T_c(0)|.
\end{equation}

If the number of MCS before verifying the histogram for flatness is
not too large, then during the simulations with each modification factor the 
checking parameter $\varepsilon$ is computed multiple times. If $\varepsilon$ 
remains below $10^{-4}$ up until the histogram meets the flatness criterion for 
this WL level, then one should save the density of states and the microcanonical 
averages and stop the simulations. When we adopt this criterion for halting the 
simulations, the final modification factors may be different for different runs.

Once obtained the density of states, one can calculate the canonical average of
any thermodynamic variable $X$ as
\begin{equation}\label{mean}
\langle X\rangle_T=\dfrac{\sum_E \langle X\rangle_E g(E) e^{-\beta E}}{\sum_E g(E) e^{-\beta E}} ,
\end{equation}
where $\langle X\rangle_E$ is the microcanonical average accumulated during the
simulations and $\beta=1/k_BT$, where $T$ is the absolute temperature measured
in units of $J/k_B$ and $k_B$ is the Boltzman's constant.

It was also observed in Ref.\cite{caparica3} that two independent similar
finite-size scaling procedures can lead to quite different results for the
exponents and the critical temperature, which in general do not agree within the 
error bars. The way to circumvent this problem is to carry out 10 
independent sets of finite-size scaling simulations. In this work, for 
each of these sets we perform simulations for $L=32,40,44,52,56,64,76,$ and 
$80$ with $n=24,20,20,16,16,16,12,$ and $12$ independent runs for each size, 
respectively. The final resulting values for the critical temperature and the  
critical exponents are obtained as an average over all sets.

\section{Finite-size scaling}

From the definition of the free energy, according to finite-size scaling theory 
\cite{fisher1,fisher2,barber}, one can obtain the zero field scaling expressions
for the magnetization, susceptibility, and specific heat, respectively, by

\begin{equation}\label{exp1}
 m\approx L^{-\beta/\nu}\mathcal{M}(tL^{1/\nu}),
\end{equation}

\begin{equation}\label{exp2}
\chi \approx L^{\gamma/\nu}\mathcal{X}(tL^{1/\nu}).
\end{equation}

\begin{equation}\label{exp3}
c \approx c_\infty + L^{\alpha/\nu}\mathcal{C}(tL^{1/\nu}),
\end{equation}
where $t=(T_c-T)/T_c$ is the reduced temperature, and $\alpha$, $\beta$, and
$\gamma$ are static critical exponents which should satisfy the scaling
relation \cite{privman}

\begin{equation}\label{scaling_relation}
 2-\alpha=d\nu=2\beta+\gamma.
\end{equation}

According to Refs. \cite{chen1993,double} we can define a set of thermodynamic
quantities defined as functions of logarithmic derivatives of the magnetization:

\begin{spreadlines}{0.9em}
\begin{align}
V_1 & \equiv 4[m^3]-3[m^4],  \label{v1} \\
V_2 & \equiv 2[m^2]-[m^4],   \label{v2} \\
V_3 & \equiv 3[m^2]-2[m^3],  \label{v3} \\
V_4 & \equiv (4[m]-[m^4])/3, \label{v4} \\
V_5 & \equiv (3[m]-[m^3])/2, \label{v5} \\
V_6 & \equiv 2[m]-[m^2],     \label{v6}
\end{align}
\end{spreadlines}
where

\begin{equation}\label{mn}
[m^n] \equiv \ln \frac{\partial \langle m^n \rangle}{\partial T}.
\end{equation}

Using these thermodynamic quantities one can determine
the static critical exponent $\nu$ , even not having yet
an estimate for the critical temperature, since taking into account Eq. 
\eqref{exp1} it is simple to show that

\begin{equation}\label{vj}
V_j\approx \frac{1}{\nu}\ln L+\mathcal{V}_j(tL^{1/\nu})
\end{equation}
for $j=1,2,...,6.$ At the critical temperature $T_c$ ($t=0$) the $\mathcal{V}_j$
are constants not dependent of the system size $L$. Having an estimate for 
the critical exponent $\nu$ it is possible to determine the critical 
temperature. According to Eqs. \eqref{exp2} and \eqref{exp3} locations of the 
extrema of the specific heat and the susceptibility, Eqs. \eqref{cv} and 
\eqref{ki}, vary asymptotically as \begin{equation}\label{tc}
 T_c(L) \approx T_c+a_qL^{-1/\nu},
\end{equation}
where $a_q$ is a quantity-dependent constant, enabling then the determination 
of $T_c$. And then, with the exponent $\nu$ and $T_c$ at hand, we can calculate
the exponents $\beta$ and $\gamma$ by the slopes of the log-log plots of Eqs. 
\eqref{exp1} and \eqref{exp2} evaluated at the critical temperature $T_c$.

\section{Results}

\subsection{The critical exponents and the critical temperature}

We carried out the simulations adopting the MCS for updating the density of 
states and the microcanonical averages were accumulated beginning from $f_7$.
The jobs were halted using the checking parameter $\varepsilon$.

\begin{figure}[!h]
\begin{center}
  \includegraphics[width=.83\linewidth, angle=-90]{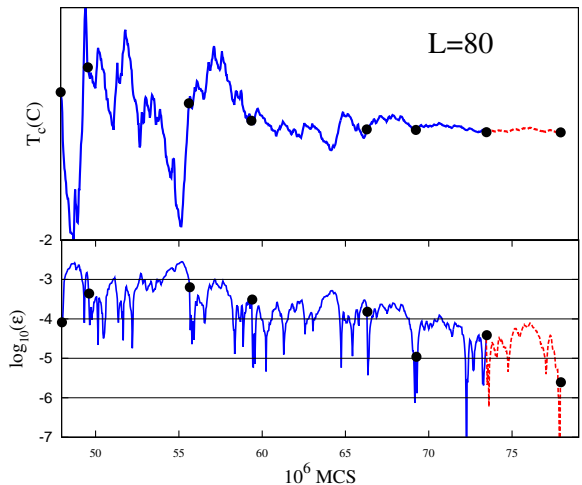}
 \end{center}
\vspace{-.5cm}
\caption{(color online). Top: Evolution of the temperature of the
maximum of the specific heat during the WL sampling, beginning from $f_{9}$ for 
a single run. The dots show where the modification factor was updated. 
Bottom: Evolution of the logarithm of the checking parameter $\varepsilon$ 
during the same simulation.} \label{checking}
\end{figure}

In Fig.\ref{checking} we show the dependence of the temperature of the maximum 
of the specific heat during the WL sampling with the number of MC steps (MCS), 
beginning from $f_9$ for a single run with $L=80$ and the evolution of 
log$_{10}(\varepsilon)$ during the same simulation. At the last WL level the 
logarithm of $\varepsilon$ remains below $-4$ indicating that the simulation can 
be halted at the end of $f_{15}$.

\begin{figure}[!ht]
\begin{center}
 \includegraphics[width=.70\linewidth, angle=-90]{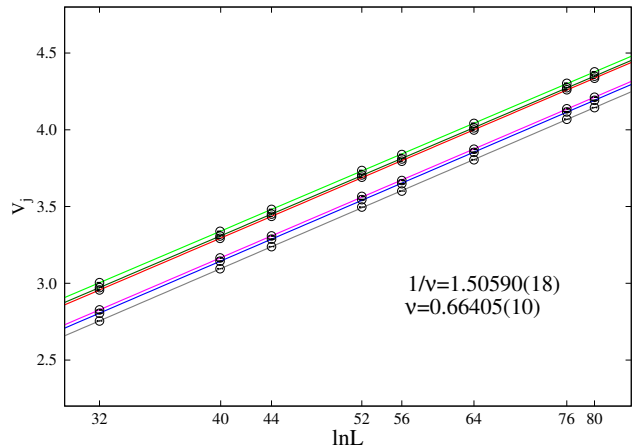}
\end{center}
\vspace{-.5cm}
\caption{(color online) Size dependence of the maxima of $V_j$. The 
slopes yield $1/\nu$.}
\label{fig:vj}
\end{figure}

\begin{figure}[!hb]
\begin{center}
 \includegraphics[width= .70\linewidth, angle=-90]{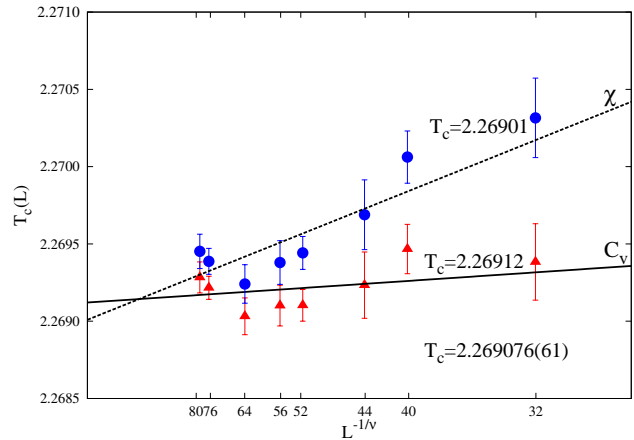}
\end{center}
\vspace{-.5cm}
\caption{(color online) Size dependence of the locations of the extrema in 
the specific heat and the susceptibility with $\nu=0.66341$.}
\label{t_c}
\end{figure}

\begin{figure}[!ht]
\begin{center}
 \includegraphics[width=.70\linewidth, angle=-90]{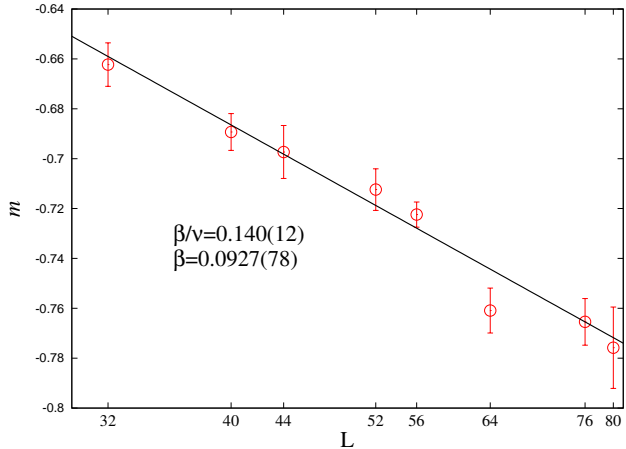}
\end{center}
\vspace{-.5cm}
\caption{(color online) Log-log plot of the size dependence of the 
finite-lattice magnetization at $T_c=2.269194$.}
\label{beta}
\end{figure}

By locating the extrema of the thermodynamic functions
defined in Eqs. \eqref{v1}-\eqref{v6} we can determine the critical exponent
$\nu$ as the inverse of the slopes of the straight lines given
by Eq. \eqref{vj}, since at $T_c$ ($t=0$) the $\mathcal{V}_j$ should be 
constants independent of the system size $L$. In Fig. \ref{fig:vj} we present this
set of lines. For each of these six slopes we calculate $\nu=1/(\frac{1}{\nu})$
with $\varDelta\nu=\varDelta(\frac{1}{\nu})/(\frac{1}{\nu})^2$ and take an
average with unequal uncertainties \cite{wong} over them. From the linear fits 
to these points we estimate $\frac{1}{\nu}=1.50590(18)$, yielding 
${\nu}=0.66405(10)$. Notwithstanding these values represent the result of 
only one of the $10$ sets of finite-size scaling simulations which were 
performed. Initially we run over all folders calculating $\nu$ in order to 
determine this exponent to the best precision. As in Refs. \cite{potts,dsde} to 
calculate the mean value of these ten results we adopted a single averaging 
instead of an average with unequal uncertainties, since the values fall into a 
Gaussian distribution and in most cases they do not agree within the error bars. 
Thus,performing the second procedure would lead to an unrealistic small error 
bar. In the first column of Table \ref{table1} we display the results for this 
exponent in each set and the final average at the last line yielding 
$\nu=0.66341(47)$. An additional discussion on the way we are taking these
averages could be held taking into account Ref. \cite{weigel}. In this work 
the authors propose a formulation using the cross correlations, as is the 
situation of our six estimates for calculating the exponent $\nu$. They assert 
that the resulting average with unequal uncertainties remains a valid estimator 
and only the error bars would be modified by the new formulation. But in our 
case the error bars obtained in each set are neglected when we take the final 
average. Moreover in the references of this work they include the following 
comment: ``Note that when using the Wang-Landau method as a direct
estimate of the density of states to be used for computing thermal expectation 
values, due to the non-Markovian nature of the algorithm there is currently no 
known approach of reliably estimating the present statistical fluctuations 
apart from repeating the whole calculation a certain number of times.''

\begin{figure}[!h]
\begin{center}
 \includegraphics[width=.70\linewidth, angle=-90]{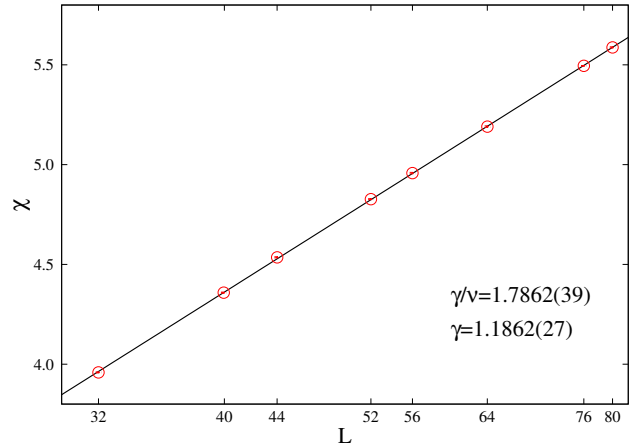}
\end{center}
\vspace{-.5cm}
\caption{(color online) Log-log plot of the size dependence of the 
finite-lattice susceptibility at $T_c=2.269194$.}
\label{gama}
\end{figure}

With the critical exponent $\nu$ accurately determined we can use Eq. \eqref{tc} 
to determine $T_c$ as the extrapolation to $L\rightarrow\infty$ 
($L^{-1/\nu}=0$) of the linear fits given by the locations of the maxima of the 
specific heat and the susceptibility defined by Eqs. \eqref{cv} and \eqref{ki}. 
In Fig. \ref{t_c} we show these linear fits that converge to $T_c$ at 
$L^{-1/\nu}=0$. At first glance the pattern seems to be disappointing with the 
points badly aligned and with large error bars, but in fact this is because all 
the results are unusually close to the correct value. The final numerical 
extrapolations are therefore excellent. In the second column of Table 
\ref{table1} we display the results for the critical temperature obtained in 
each set and at the last line the mean value giving $T_c=2.269194(45)$.

\begin{table}[htb]
   \centering
   \setlength{\arrayrulewidth}{2\arrayrulewidth}  
   \setlength{\belowcaptionskip}{10pt} 
\begin{tabular}{c|c|c|c}
\hline
  $\nu$ & $T_c$ & $\beta$ & $\gamma$ \\
\hline  
\hspace{.2cm}0.66405(20)\hspace{.2cm}&\hspace{.2cm}2.269076(86)\hspace{.2cm}
 
&\hspace{.2cm}0.0926(78)\hspace{.2cm}&\hspace{.2cm}1.1850(34)\hspace{.2cm}\\
\hspace{.2cm}0.66553(64)\hspace{.2cm}&\hspace{.2cm}2.26938(10)\hspace{.2cm}
&\hspace{.2cm}0.0690(95)\hspace{.2cm}&\hspace{.2cm}1.1762(38)\hspace{.2cm}\\ 
   
\hspace{.2cm}0.66242(67)\hspace{.2cm}&\hspace{.2cm}2.269189(85)\hspace{.2cm}
&\hspace{.2cm}0.0821(72)\hspace{.2cm}&\hspace{.2cm}1.1822(39)\hspace{.2cm}\\ 
  
\hspace{.2cm}0.66150(41)\hspace{.2cm}&\hspace{.2cm}2.269162(75)\hspace{.2cm}
&\hspace{.2cm}0.0816(73)&\hspace{.2cm}1.1835(28)\hspace{.2cm}\\               
  
\hspace{.2cm}0.66139(42)\hspace{.2cm}&\hspace{.2cm}2.26899(11)\hspace{.2cm}
&\hspace{.2cm}0.093(10)\hspace{.2cm}&\hspace{.2cm}1.1851(31)\hspace{.2cm}\\  
   
\hspace{.2cm}0.66276(56)\hspace{.2cm}&\hspace{.2cm}2.26924(13)\hspace{.2cm}
&\hspace{.2cm}0.083(11)\hspace{.2cm}&\hspace{.2cm}1.1845(40)\hspace{.2cm}\\  
   
\hspace{.2cm}0.66353(43)\hspace{.2cm}&\hspace{.2cm}2.269329(75)\hspace{.2cm}
&\hspace{.2cm}0.0705(71)\hspace{.2cm}&\hspace{.2cm}1.1795(26)\hspace{.2cm}\\  
   
\hspace{.2cm}0.66498(63)\hspace{.2cm}&\hspace{.2cm}2.26905(11)\hspace{.2cm}
&\hspace{.2cm}0.0905(90)\hspace{.2cm}&\hspace{.2cm}1.1819(34)\hspace{.2cm}\\ 
  
\hspace{.2cm}0.66522(68)\hspace{.2cm}&\hspace{.2cm}2.26912(11)\hspace{.2cm}
&\hspace{.2cm}0.0867(89)\hspace{.2cm}&\hspace{.2cm}1.1808(36)\hspace{.2cm}\\ 
   
\hspace{.2cm}0.66271(47)\hspace{.2cm}&\hspace{.2cm}2.269410(99)\hspace{.2cm}
&\hspace{.2cm}0.0694(99)\hspace{.2cm}&\hspace{.2cm}1.1806(41)\hspace{.2cm}\\
\hline
\hline
\hspace{.2cm}0.66341(47)\hspace{.2cm} 
&\hspace{.2cm}2.269194(45)\hspace{.2cm}
&\hspace{.2cm}0.0818(30)\hspace{.2cm}&\hspace{.2cm}1.1819(17)\hspace{.2cm}
\\
\hline
\end{tabular}
\caption{Ten finite size scaling results for the critical temperature 
$T_c$, and the exponents $\nu$, $\beta$ and $\gamma$. The average over all runs 
are shown at the last line.}
\label{table1}
\end{table}

\begin{table*}[!htb]
\begin{ruledtabular}
\begin{tabular}{l c c c c}
Method                                  & $\alpha$     & $\beta$        & 
$\gamma$       & $\nu$          \\
\hline
\multicolumn{5}{c}{$q=4$ Potts model} \\ \hline
Conjectured values \cite{Wu1982}         & $\frac{2}{3}\cong0.667$& 
$\frac{1}{12}\cong0.083$ & $\frac{7}{6}\cong1.167$  & $\frac{2}{3}\cong0.667$  
\\
Kadanoff variational RG \cite{dasgupta} & $0.488$      & $0.091$        & 
$1.330$        & $0.756$        \\
Duality invariant RG \cite{hu}  & $0.4870$     & $-$ & $-$  & $0.7565$ \\
Critical dynamics \cite{alternative} & $-$ & $0.0835(4)$ & $-$  & $0.669(6)$ \\
Entropic sampling \cite{potts}     & $0.5084(48)$ & $0.0877(37)$  
 & $1.3161(69)$   & $0.7076(10)$   \\    
 \hline
\multicolumn{5}{c}{Baxter-Wu model} \\ \hline
Exact solution \cite{baxter}         & $\frac{2}{3}\cong0.667$& 
$\frac{1}{12}\cong0.083$ & $\frac{7}{6}\cong1.167$  & $\frac{2}{3}\cong0.667$ \\
Critical dynamics \cite{wagner} & $-$ & $0.0817(23)$ & $-$  & $0.621(9)$ \\
Critical dynamics \cite{everaldo} & $-$ & $0.080(2)$ & $-$  & $0.67(1)$ \\
This work  & $0.6545(77)$ & $0.0818(30)$  & $1.1819(17)$   & $0.66341(47)$ \\
\end{tabular}
\end{ruledtabular}
\caption{Estimates of $\alpha$, $\beta$, $\gamma$, and $\nu$ compared to 
results 
obtained with other techniques, conjectured, and exact values.}
\label{table2}
\end{table*}

Next, with the critical temperature $T_c$ and the critical exponent $\nu$ 
estimated to a high precision, we can use Eqs.\eqref{exp1}-\eqref{exp2} to 
determine the exponents $\frac{\beta}{\nu}$ and $\frac{\gamma}{\nu}$ by the 
slopes of the log-log plots. Fig. \ref{beta} and Fig. \ref{gama} show 
this finite-size scaling behavior for each exponent, yielding  
$\frac{\beta}{\nu}=0.140(12)$ and $\frac{\gamma}{\nu}=1.7862(39)$,
respectively. We then estimate $\beta=\nu\frac{\beta}{\nu}$ with
$\varDelta\beta= \frac{\beta}{\nu}\varDelta\nu+\nu\varDelta\frac{\beta}{\nu}$,
and similarly for $\gamma$ and $\varDelta\gamma$, obtaining $\beta=0.0926(78)$,
and $\gamma=1.1850(34)$. Over again, these values are calculated at the first 
set. In the two last columns of Table \ref{table1} we show the results for all 
folders and the best estimates at the last line, obtaining $\beta=0.0818(30)$, 
and $\gamma=1.1819(17)$. As a final result, using the scaling relation given by 
Eq. \eqref{scaling_relation} we can calculate the exponent 
$\alpha=2-2\beta-\gamma$ with $\varDelta\alpha=2\varDelta\beta+\varDelta\gamma$ 
giving $\alpha=0.6545(77)$.

In Table \ref{table2} we compare our final estimates 
of the critical exponents to other well-established values. It is noteworthy 
the proximity of our estimates of the critical exponents with the 
exact values. The good resolution we have obtained in this study 
corroborates the strength of the adopted technique.

\subsection{The $a$ and $b$ coefficients of the 
 divergence of the microcanonical inverse temperature}

The microcanonical inverse temperature is defined as

\begin{equation}\label{derivative}
\frac{1}{T}=\frac{\partial S}{\partial E}=\lim_{\varDelta E 
\to 0}\frac{\varDelta S}{\varDelta E}.
\end{equation}

In the ground state this limit should be infinity. Nevertheless for any 
discrete model, as in the Baxter-Wu model, $\varDelta E=const$.
Therefore the limit becomes exact only if $L\rightarrow\infty$ 
($E_{min}\rightarrow-\infty$), where $L$ is the linear lattice size \cite{dsde}.

\begin{figure}[!hb]\centering
\begin{center}
 \includegraphics[width=.69\linewidth,angle=-90]{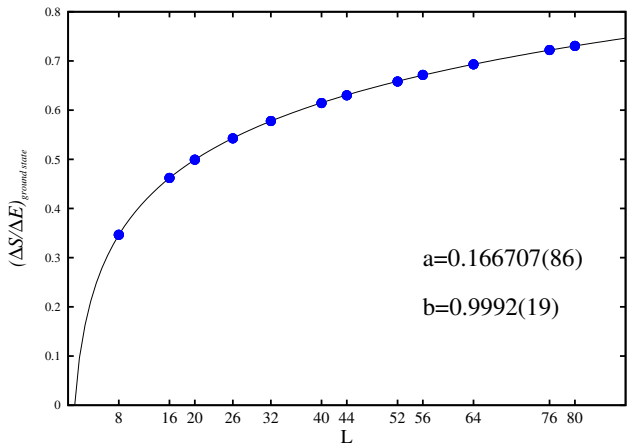}
\end{center}
\caption{(Color online) Size dependence of $\frac{\varDelta S}{\varDelta E}$ 
at the ground state of the first set of simulations. The line is the best 
fitting curve of these data to $a\ln(bL)$. The error bars are smaller than the 
symbols.}
\label{dsde}
\end{figure}

\begin{table}[!hb]
   \centering
   \setlength{\arrayrulewidth}{2\arrayrulewidth}  
   \setlength{\belowcaptionskip}{10pt} 
\begin{tabular}{c|c}
\hline
\hspace{.5cm} $a$\hspace{1cm} &\hspace{1cm} $b$\hspace{.5cm}  \\
\hline
\hspace{.5cm} 0.16680(10)\hspace{1cm} &\hspace{1cm} 0.9972(16)\hspace{.5cm}  \\
\hspace{.5cm} 0.16714(12)\hspace{1cm} &\hspace{1cm} 0.9897(25)\hspace{.5cm}  \\
\hspace{.5cm} 0.16692(10)\hspace{1cm} &\hspace{1cm} 0.9946(20)\hspace{.5cm}  \\
\hspace{.5cm} 0.16701(11)\hspace{1cm} &\hspace{1cm} 0.9928(23)\hspace{.5cm}  \\
\hspace{.5cm} 0.16659(16)\hspace{1cm} &\hspace{1cm} 1.0028(34)\hspace{.5cm}  \\
\hspace{.5cm} 0.16663(11)\hspace{1cm} &\hspace{1cm} 1.0006(24)\hspace{.5cm}  \\
\hspace{.5cm} 0.16621(10)\hspace{1cm} &\hspace{1cm} 1.0100(21)\hspace{.5cm}  \\
\hspace{.5cm} 0.16669(11)\hspace{1cm} &\hspace{1cm} 0.9988(25)\hspace{.5cm}  \\
\hspace{.5cm} 0.16646(10)\hspace{1cm} &\hspace{1cm} 1.0052(20)\hspace{.5cm}  \\
\hspace{.5cm} 0.16663(10)\hspace{1cm} &\hspace{1cm} 1.0001(20)\hspace{.5cm}  \\
\hline \hline
\hspace{.5cm} 0.166707(86)\hspace{1cm} &\hspace{1cm} 0.9992(19)\hspace{.5cm} \\
\hline 
\end{tabular}
\caption{Estimates by entropic sampling simulations of the parameters $a$ and 
$b$ for the Baxter-Wu model.}
\label{table3}
\end{table}

In the Baxter-Wu model the ground state has four configurations, one with
all spins positive and three more where two sub-lattices are negated 
successively. If a single spin is flipped, it looses six negative 
links from the neighboring triangular faces and gains six positive ones. 
The energy gap between the ground state and the next state is therefore 
$\varDelta E=6-(-6)=12$, where we take the coupling constant $J$ as 1. The 
first higher level has $4L^2$ configurations yielding $\varDelta S=\ln4+2\ln 
L-\ln4$ and, since $\varDelta E=12$, we have
\begin{equation}\label{derivative_i2}
\frac{\varDelta S}{\varDelta E}=\frac{1}{6}\ln L\cong0.1667\ln L.
\end{equation}
For notational simplicity we set $k_B=1$. Just as in the cases discussed in Ref.
\cite{dsde} the inverse temperature of the Baxter-Wu model diverges in the 
ground state as $a\ln (bL)$, yielding $a=\frac{1}{6}$ and $b=1$. For estimating 
the coefficients $a$ and $b$ from our simulational data we added four smaller 
sizes, $L=8,16,20,26$. In Fig. \ref{dsde} we show the dependence of 
$\frac{\varDelta S}{\varDelta E}$ at the ground state on the lattice sizes using 
the outcomes of the first set of simulations, along with the best fitting curve 
to $a\ln(bL)$. The agreement is excellent. In Table \ref{table3} we display the 
values obtained in our entropic simulations for $a$ and $b$. The ten initial 
lines correspond to results obtained in each set and the last line is a single 
average over all sets neglecting the error bars.  One can see that we get very 
accurate results for $a$ and $b$, which agree within error bars equal to 
$\pm1\sigma$ with the exact coefficients $a=\frac{1}{6}$ and $b=1$. 

\section{Conclusions}

We carried out a high-resolution study of the static critical behavior of the 
Baxter-Wu model using a refined entropic sampling procedure based on the 
Wang-Landau method. Our results present an impressive agreement with the 
exact values for the critical exponents and with the critical temperature 
as well. Such upshot corroborates the correctness of the improvements suggested
to the original Wang-Landau sampling. Finally we calculated the exact values 
for the recently proposed coefficients $a$ and $b$ for the divergence of the 
microcanonical inverse temperature in the ground state and we also obtained an
excellent concordance of the simulation estimates with them. 

\section{Acknowledgment}

We acknowledge the computer resources provided by LCC-UFG and IF-UFMT. L. N. 
Jorge acknowledges the support by FAPEG.


\begin{thebibliography}{99}

\bibitem{bw1}R. J. Baxter and F.Y. Wu, Phys. Rev. Lett. \textbf{31}, 1294 
(1973).

\bibitem{bw2}R. J. Baxter and F.Y. Wu, Aust. J. Phys. \textbf{27}, 357 (1974).

\bibitem{bw3}R. J. Baxter, Aust. J. Phys. \textbf{27}, 368 (1974).

\bibitem{berg1}B. A. Berg and T. Neuhaus, Phys. Rev. Lett. \textbf{68}, 9 
(1992).

\bibitem{berg2}B. A. Berg and T. Celik, Phys. Rev. Lett. \textbf{69}, 2292 
(1992).

\bibitem{berg3}B. A. Berg, U. Hansmann, and T. Neuhaus, Phys. Rev. B 
\textbf{47}, 497 (1993).

\bibitem{janke}W. Janke and S. Kappler, Phys. Rev. Lett. \textbf{74}, 212  1995.

\bibitem{lee}J. Lee, Phys. Rev. Lett. \textbf{71}, 211 (1993).

\bibitem{broad1}P. M. C. Oliveira, T. J. P. Penna, and H. J. Herrmann, Braz. J. 
Phys. \textbf{26}, 677 (1996).

\bibitem{broad2}P. M. C. Oliveira, T. J. P. Penna, and H. J. Herrmann, Eur. 
Phys. J. B \textbf{1}, 205 (1998).

\bibitem{broad3}P. M. C. Oliveira, Eur. Phys. J. B \textbf{6}, 111 (1998).

\bibitem{WangLandau2001}{F. Wang and D. P. Landau, Phys. Rev. Lett. {\bf 86}, 
2050 (2001); Phys. Rev. E \textbf{64}, 056101 (2001).}

\bibitem{ZhouBhatt2005}{C. Zhou and R. N. Bhatt, Phys. Rev. E {\bf 72}, 025701 
(2005).}

\bibitem{belardinelli} R. E. Belardinelli and V. D. Pereyra, J. Chem. Phys. 
\textbf{127}, 184105 (2007).

\bibitem{caparica} A.A. Caparica and A.G. Cunha-Netto, Phys. Rev. E 
\textbf{85}, 046702 (2012).

\bibitem{caparica3} A.A. Caparica, Phys. Rev. E \textbf{89}, 043301 (2014).

\bibitem{wood}Wood, D. W. and Griffiths, H.P., J. Phys. C: Solid State Phys. 
\textbf{5}, L253-5 (1972).

\bibitem{caparica1}L.S. Ferreira and A.A. Caparica, 
Int. J. Mod. Phys. C \textbf{23}, 1240012 (2012).
 
\bibitem{caparica2}L.S. Ferreira, A.A. Caparica, M. A. Neto, and M. D. 
Galiceanu, J. Stat. Mech., \textbf{2012}, P10028 (2012).

\bibitem{mcs} A Monte Carlo sweep consists of $L^2$ spin-flip trials in the 2D 
lattice.
 
 \bibitem{Wu1982}{F.Y Wu, The Potts model, Rev. Mod. Phys. {\bf 54}, 235 
(1982).}

 \bibitem{fisher1} M.E. Fisher, in \textit{Critical Phenomena}, edited by M. S. 
Green (Academic, New York, 1971).
 
 \bibitem{fisher2} M.E. Fisher and M.N. Barber, Phys. Rev. Lett. \textbf{28}, 
1516 (1972).
 
 \bibitem{barber} \textit{Phase Transitions and Critical Phenomena}, edited by 
C. Domb and J. L. Lebowitz
 (Academic, New York, 1974), Vol. 8.
 
 \bibitem{privman} V. Privman, P.C. Hohenberg, and A. Aharony, in \textit{Phase 
Transitions and Critical Phenomena},
 eduted by C. Domb and J. L. Lebowitz (Academic, New York, 1991), Vol. 14, p. 1.
 
 \bibitem{chen1993} K. Chen,A.M. Ferrenberg, and D.P. Landau, Phys. Rev. 
B \textbf{48}, 3249 (1993).
 
\bibitem{double}A.A. Caparica, A. Bunker, and D.P. Landau, Phys. Rev. B 
\textbf{62}, 9458 (2000); There is a misprinting
 in Eq.(3) in this paper, which should be $V_5\equiv (3[m]-[m^3])/2$.
 
\bibitem{wong}S.S.M. Wong, \textit{Computational Methods in Physics and 
Engineering}, 2\textit{nd} edition, World Scientific Publishing Co. Pte. Ltd. 
(1997).

\bibitem{alternative} H. A. Fernandes, E. Arashiro, J. R. Drugowich 
de Fel\'icio, 
and A. A. Caparica, Physica A \textbf{366} 255–264 (2006).

\bibitem{potts} A. A. Caparica, S. A. Le\~ao, and C. J. DaSilva, Physica A 
\textbf{438} 447–453 (2015).

\bibitem{dsde} A. A. Caparica, S. A. Le\~ao, and C. J.DaSilva, Braz. J. Phys., 
DOI 10.1007/s13538-015-0361-8 (2015).

\bibitem{weigel} M. Weigel and W. Janke, Phys. Rev. E \textbf{81}, 066701 
(2010).

\bibitem{dasgupta} C. Dasgupta, Phys. Rev. B {\bf 15}, 3460 (1977).
 
\bibitem{hu} B. Hu, J. Phys. A {\bf 13}, L321 (1980).

\bibitem{baxter} R. J. Baxter, \textit{Exactly Solved Models in Statistical 
Mechanics}, p. 320, (Academic Press, London, 1982).

\bibitem{wagner} M. Santos and W. Figueiredo, Phys. Rev. E \textbf{63}, 042101 
(2001).

\bibitem{everaldo} E. Arashiro and J. R. Drugowich de Fel\'icio, Phys. Rev. E 
\textbf{67}, 046123 (2003).

\end{thebibliography}
\end{document}